\pdfoutput=1
\documentclass[sigconf]{acmart}

\copyrightyear{2027}
\acmYear{2027}
\setcopyright{cc}
\setcctype{by}
\acmConference[ASPDAC '27]{32nd Asia and South Pacific Design Automation Conference}{January 25--28, 2027}{Tokyo, Japan}
\acmBooktitle{32nd Asia and South Pacific Design Automation Conference (ASPDAC '27), January 25--28, 2027, Tokyo, Japan}
\acmDOI{10.1145/3840404.3848175}
\acmISBN{979-8-4007-2923-2/2027/01}

\usepackage{textcomp}
\usepackage{xspace}
\usepackage{tikz}

\usetikzlibrary{arrows.meta,positioning,fit,calc,backgrounds}
\usepackage{url}

\AtBeginDocument{%
  \providecommand\BibTeX{{%
    \normalfont B\kern-0.5em{\scshape i\kern-0.25em b}\kern-0.8em\TeX}}}

\mathchardef\UrlBigBreakPenalty=10000

\newcommand{\sysname}{\textsc{Ares}\xspace}
\DeclareRobustCommand{\cbadge}[1]{\tikz[baseline=(b.base)]%
  \node[circle,fill=cyan!45!black,text=white,font=\bfseries\scriptsize,
        inner sep=1pt,minimum size=9pt](b){#1};}
\definecolor{badgeorange}{HTML}{D95F02}
\DeclareRobustCommand{\cbadgeorange}[1]{\tikz[baseline=(b.base)]%
  \node[circle,fill=badgeorange,text=white,font=\bfseries\scriptsize,
        inner sep=1pt,minimum size=9pt](b){#1};}

\begin{document}

\title{\sysname{}: Adaptive Reasoning-Effort Steering for PPA- and Cost-Aware RTL
Optimization with LLM Agents}

\author{Stef Cuyckens}
\email{stef.cuyckens@kuleuven.be}
\affiliation{%
  \institution{KU Leuven}
  \city{Leuven}
  \country{Belgium}}

\author{Mihaela Jivanescu}
\email{mihaela.jivanescu@nokia-bell-labs.com}
\affiliation{%
  \institution{Nokia Bell Labs}
  \city{Antwerp}
  \country{Belgium}}

\author{Jun Yin}
\email{jun.yin@kuleuven.be}
\affiliation{%
  \institution{KU Leuven}
  \city{Leuven}
  \country{Belgium}}

\author{Chao Fang}
\authornote{Corresponding author.}
\email{chao.fang@kuleuven.be}
\affiliation{%
  \institution{KU Leuven}
  \city{Leuven}
  \country{Belgium}}

\author{Marian Verhelst}
\email{marian.verhelst@kuleuven.be}
\affiliation{%
  \institution{KU Leuven}
  \city{Leuven}
  \country{Belgium}}

\renewcommand{\shortauthors}{Cuyckens et al.}

\begin{abstract}
Large language model (LLM) agents optimize the power, performance, and area
(PPA) of register-transfer-level (RTL) designs by iterating over edits, synthesis,
and PPA analysis, paying a dollar cost for every LLM call.
{Prior agents report the quality reached without its normalized cost, attribute
that quality to an engineered cross-design memory, and hold the reasoning effort
of every call fixed.
We propose \sysname{} with three corresponding innovations.
(1) We introduce a normalized dollar cost per LLM call reported alongside the
figure of merit (FoM), enabling fair comparison across effort levels and optimizers.
(2) Using this accounting, we find the construction of the long-term memory
matters little.
An engineered memory brings no dependable gain over a plain concatenation of
the same experience.
(3) We instead adapt the per-call reasoning effort by escalating to deeper
reasoning only once progress at a lower effort stalls, via a patience counter
fit on 21 training designs, allocating reasoning where it pays rather than
uniformly across all iterations.}
On three test designs unseen during training, the effort policy lowers the FoM by
22--28\% with Opus 4.6 and by 18\nobreakdash--34\% with GPT-5.5, where the best fixed effort
reaches 20--24\% and 19\nobreakdash--29\%, respectively, at equal normalized cost.
\sysname{} closes on average 91\% of the gap from an LLM-drafted multiply-accumulate
unit to its highly hand-optimized counterpart. It also reaches a 5\%
lower FoM than state-of-the-art Dr.\,RTL at only 35\% of
its cost.
\end{abstract}



\maketitle

\section{Introduction}
\label{sec:intro}

Meeting power, performance, and area (PPA) targets is a central challenge of
digital hardware design, and improving the register-transfer-level (RTL)
description that determines them has traditionally required scarce human expert
effort~\cite{rtlopt,chipnemo,surveyeda}. Much RTL therefore ships functionally correct
but unoptimized, whether drafted quickly by an engineer or generated by an
LLM.

Automating the
optimization of such RTL with large-language-model (LLM) agents has recently drawn
increased
attention~\cite{rtlopt,pluto,surveyeda}. An LLM agent pairs an LLM with
electronic design automation (EDA) tools. It takes an existing, unoptimized design
as input, edits it, synthesizes the edit, reads back the resulting PPA, and
repeats. If the new design has a lower figure of merit (FoM), for example a
power-area-delay product defined against the original design, it is retained. The
quality of these edits
improves with the capability of the LLM. Capable agents therefore rely on large commercial
models~\cite{budgetmlagent}, whose rapidly growing token bills~\cite{tokenecon}
already put a single agentic run at several dollars~\cite{morewithless}. The
cost of an optimization run thus deserves as much attention as the FoM it reaches.
Prior art falls short on both counts. Some agents that optimize existing RTL
score only area and timing, leaving power out of the
objective~\cite{drrtl,cktevo}. On cost, every state-of-the-art optimizer falls
short: at best it reports one aggregate dollar figure for a whole study, {and none provides 
a unified metric that relates optimization quality to its dollar cost, allowing fair comparison 
across optimizers~\cite{drrtl,cktevo,poet}.}

\begin{figure*}[t!]
\centering
\resizebox{0.97\textwidth}{!}{%
\begin{tikzpicture}[
  font=\footnotesize,
  >={Stealth[round]},
  line/.style={line width=0.9pt,rounded corners=3pt},
  box/.style={rounded corners=2.5pt,draw,line width=0.8pt,align=center,inner sep=2.5pt,
              minimum height=8.5mm,minimum width=24mm},
  design/.style={box,fill=yellow!12,draw=olive!45},
  input/.style={box,fill=yellow!35,draw=olive!80!black,line width=1pt},
  agent/.style={box,fill=violet!8,draw=violet!40},
  verify/.style={box,fill=orange!9,draw=orange!50},
  synth/.style={box,fill=blue!5,draw=blue!35},
  fom/.style={box,fill=green!18,draw=green!55!black,line width=1pt},
  ctrl/.style={box,fill=cyan!15,draw=cyan!72!black,line width=1pt},
  mem/.style={rounded corners=2.5pt,draw,line width=0.8pt,align=center,inner sep=2.5pt,
             fill=teal!7,draw=teal!40,minimum width=24mm,minimum height=8.5mm},
  badge/.style={circle,fill=cyan!45!black,text=white,font=\bfseries\scriptsize,
                inner sep=1pt,minimum size=9pt},
  lbl/.style={font=\scriptsize},
  slbl/.style={font=\scriptsize,text=black!75},
  flbl/.style={font=\scriptsize\itshape},
  okarr/.style={line,->,black!60},
  mainarr/.style={line width=1.7pt,rounded corners=3pt,->,black!60},
  failarr/.style={line,->,orange!80!black!70,dashed},
  ]

  \node[input] (v0)   at (0,2.5) {unoptimized RTL $v_0$};
  \node[design] (best) at (0,1.0) {best design yet\\\scriptsize current optimum};
  \node[input] (tb)   at (0,-0.55) {testbench\\\scriptsize generated from $v_0$};
  \draw[okarr] (v0) -- (best) node[lbl,midway,right=1pt]{initialize};
  \begin{scope}[on background layer]
    \node[draw=olive!45,dashed,rounded corners=3pt,inner sep=3.5pt,
          fill=yellow!4,fit=(v0)(best)(tb)] (dgroup) {};
  \end{scope}
  \node[above=2pt of dgroup.north,text=olive!70!black,font=\small\bfseries]
        {Input design $\boldsymbol{v_0}$};

  \node[align=center,font=\footnotesize,inner sep=1pt] (llmtxt) at (3.7,2.4) {LLM agent\\\scriptsize edits the RTL};
  \node[rounded corners=1.5pt,draw,dashed,line width=0.6pt,draw=teal!40,fill=teal!7,
        text=black!70,font=\scriptsize,align=center,inner sep=1.5pt] (stm) at (3.7,1.75)
        {short-term memory\\(current run)};
  \coordinate (llmbot) at (3.7,1.3);
  \begin{scope}[on background layer]
    \node[agent,fit=(llmtxt)(stm)(llmbot),inner sep=2pt] (llm) {};
  \end{scope}
  \node[font=\footnotesize,anchor=west] (efftitle) at (11.05,3.86)
        {\textbf{Adaptive reasoning effort}, set per call};
  \node[font=\scriptsize,text=black!75,anchor=west] (effmech) at (11.05,3.53)
        {stalled and failed iterations charge a patience counter $C$};
  \node[box,fill=orange!22,draw=orange!80!black,minimum width=19mm,minimum height=7mm]
        (medch) at (12.35,3.02) {\textbf{medium}\\\scriptsize dependable, cheap};
  \node[box,fill=violet!18,draw=violet!70!black,minimum width=19mm,minimum height=7mm]
        (highch) at (16.45,3.02) {\textbf{high}\\\scriptsize deeper, higher cost};
  \draw[line,->,black!70] ([yshift=1.6mm]medch.east) -- ([yshift=1.6mm]highch.west)
        node[font=\scriptsize,midway,fill=cyan!15,inner sep=0.8pt]{escalate: $C\!\ge\!p$};
  \draw[line,->,black!70] ([yshift=-1.6mm]highch.west) -- ([yshift=-1.6mm]medch.east)
        node[font=\scriptsize,midway,fill=cyan!15,inner sep=0.8pt] (dislbl) {discharge};
  \begin{scope}[on background layer]
    \node[ctrl,fit=(efftitle)(effmech)(medch)(highch)(dislbl)] (eff) {};
  \end{scope}
  \node[badge] at (eff.west |- efftitle.center) {3};
  \node[mem] (md) at (4.7,-0.3) {long-term memory\\\scriptsize markdown};
  \node[badge] at (md.north west) {2};
  \draw[line,->,teal!45] (md.north) -- ++(0,0.55) -| (llm.south);
  \node[font=\tiny\itshape,text=black!55,anchor=north] at (4.7,-0.78)
        {};

  \node[verify,right=9mm of llm] (fverify) {functional verify\\\scriptsize long random TB + SEC vs $v_0$};
  \draw[mainarr] (best.east) -- ++(0.45,0) |- (llm.west);
  \draw[mainarr] (llm) -- (fverify);
  \draw[failarr] (fverify.north) -- (fverify.north |- 0,3.15) -| ([xshift=7mm]llm.north);
  \node[flbl,text=orange!65!black,anchor=north] at (5.6,3.08) {fail: discard};

  \node[synth] (fmax)  at (2.6,-1.9) {synthesize\\\scriptsize max $f$ (Synopsys DC)};
  \node[synth,right=5mm of fmax]  (area)  {synthesize @ $f$\\\scriptsize area (Synopsys DC)};
  \node[verify,right=5mm of area] (gv)    {gate-level verify\\\scriptsize netlist + TB};
  \node[synth,right=5mm of gv] (pwr)  {power\\\scriptsize PrimeTime};
  \node[fom,right=6mm of pwr]     (fom)   {\textbf{Figure of Merit}\\\scriptsize $f\cdot$area$\cdot$power (norm.\ $v_0$)};
  \begin{scope}[on background layer]
    \node[draw=blue!35,dotted,rounded corners=3pt,inner sep=3.5pt,
          fill=blue!2,fit=(fmax)(area)(gv)(pwr)] (sgroup) {};
  \end{scope}
  \node[slbl,anchor=north west,text=blue!45!black!70] at ($(sgroup.south west)+(0,-1.5pt)$)
        {\textbf{synthesis \& PPA measurement (commercial flow)}};
  \draw[mainarr] (fmax) -- (area);
  \draw[mainarr] (area) -- (gv);
  \draw[mainarr] (gv) -- (pwr);
  \draw[mainarr] (pwr) -- (fom);
  \draw[mainarr] (fverify.south) -- ++(0,-0.6) -| (fmax.north)
        node[flbl,pos=0.82,left=2pt,text=black!70]{pass};
  \coordinate (tbrail) at (0,-1.15);
  \draw[line,dashed,black!60] (tb.south) -- (tbrail);
  \draw[okarr,dashed] (tbrail) -| ([xshift=5mm]fverify.south);
  \draw[okarr,dashed] (tbrail) -| ([xshift=-2mm]gv.north);
  \draw[okarr,dashed] (tbrail) -| (pwr.north);
  \node[flbl,text=black!60,anchor=south] at ($(gv.north)!0.5!(pwr.north)+(0,0.34)$)
        {random vectors};

  \coordinate (railY) at (0,-2.88);
  \draw[line,->,green!55!black,line width=1.7pt] (fom.south) |- (-1.9,-2.88) -- (-1.9,1.0) -- (best.west);
  \node[flbl,text=green!45!black] at (5.9,-3.13)
        {if FoM beats best: keep as new best; else discard; then iterate};
  \draw[failarr] (gv.south) -- (gv.south |- railY)
        node[flbl,pos=0.62,right=1.5pt,text=orange!65!black]{fail: discard};

  \draw[line,->,cyan!65!black] (eff.west) -| ([xshift=-3mm]llm.north);

  \begin{scope}[xshift=6mm]
  \begin{scope}[on background layer]
    \fill[black!2,rounded corners=3pt,draw=black!18,line width=0.6pt]
      (11.2,-1.1) rectangle (17.55,2.25);
  \end{scope}
  \node[slbl,anchor=north] at (14.4,2.2) {\textbf{Result}};
  \draw[line width=0.7pt,->] (11.8,-0.58) -- (17.3,-0.58)
        node[lbl,pos=0.97,below=1pt](costlbl){cost};
  \node[badge,left=2pt of costlbl] {1};
  \draw[line width=0.7pt,->] (11.8,-0.58) -- (11.8,1.85)
        node[lbl,above=0pt]{best FoM};
  \fill[black] (12.1,1.57) circle (1.6pt);
  \node[lbl,anchor=west] at (12.25,1.66) {unoptimized $v_0$};
  \draw[violet!70!black,dashed,line width=0.8pt]
    (12.1,1.57) .. controls (13.4,0.87) and (15.5,0.52) .. (16.9,0.42);
  \node[violet!70!black,lbl,anchor=north] at (16.05,0.38) {fixed high};
  \draw[orange!88!black,line width=1.0pt]
    (12.1,1.57) .. controls (12.8,0.97) and (13.3,0.80) .. (13.9,0.74);
  \draw[orange!88!black,dashed,line width=0.8pt]
    (13.9,0.74) .. controls (15.0,0.70) and (16.2,0.68) .. (17.1,0.67);
  \node[orange!88!black,lbl,anchor=south] at (15.6,0.74) {stuck at medium};
  \node[orange!88!black,lbl,anchor=east] at (13.3,0.55) {medium};
  \draw[violet!82!black,line width=1.5pt]
    (13.9,0.74) .. controls (14.3,0.36) and (14.55,0.19) .. (15.1,0.10);
  \fill[violet!82!black] (15.1,0.10) circle (1.6pt);
  \node[violet!82!black,lbl,align=right,anchor=east] at (14.15,0.22)
        {high\\effort};
  \draw[->,gray,line width=0.7pt] (15.6,-0.24) -- (15.22,-0.02);
  \node[lbl,anchor=south] at (15.85,-0.52) {lower FoM at lower cost};
  \end{scope}

  \draw[line,->,teal!45,dashed] (11.8,-0.3) -- (md.east);
  \node[flbl,text=teal!45!black,anchor=south] at (9.9,-0.24)
        {After optimizing, update long-term memory};
\end{tikzpicture}}
\Description{Flow diagram of the Ares optimizer. On the left are the inputs:
the unoptimized RTL v0, the best design found so far, and a testbench
generated from v0. In the center, an LLM agent with a short-term memory edits
the best RTL and reads a long-term memory in markdown, which is updated after
each run. Each edit is checked by functional verification against v0 and then
measured on a commercial flow (synthesis, gate-level verification, and power
analysis), which gives a figure of merit that combines delay, area, and power,
normalized to v0. An edit is kept only if its figure of merit beats the best
one. At the top right, the adaptive reasoning effort switches each call between
medium and high effort with a patience counter. A small sketch at the right
plots the best figure of merit against cost: fixed medium effort stalls, fixed
high effort costs more, and the adaptive policy reaches a lower figure of merit
at a lower cost. Badges 1 to 3 mark the three contributions.}
\caption{\textbf{Overview of \sysname.} The LLM agent edits the running-best
RTL; each candidate is verified against the input design $v_0$, synthesized on the
commercial flow, and kept only if it lowers the FoM.
{Our contributions: \cbadge{1} the per-call cost metric enabling proper comparison,
\cbadge{2} the finding that the construction of the cross-design long-term memory matters little, and \cbadge{3} an adaptive reasoning-effort policy that escalates once progress stalls. On the right, our adaptive effort reaches a lower FoM than fixed
effort at the same cost. %
}}
\label{fig:teaser}
\end{figure*}
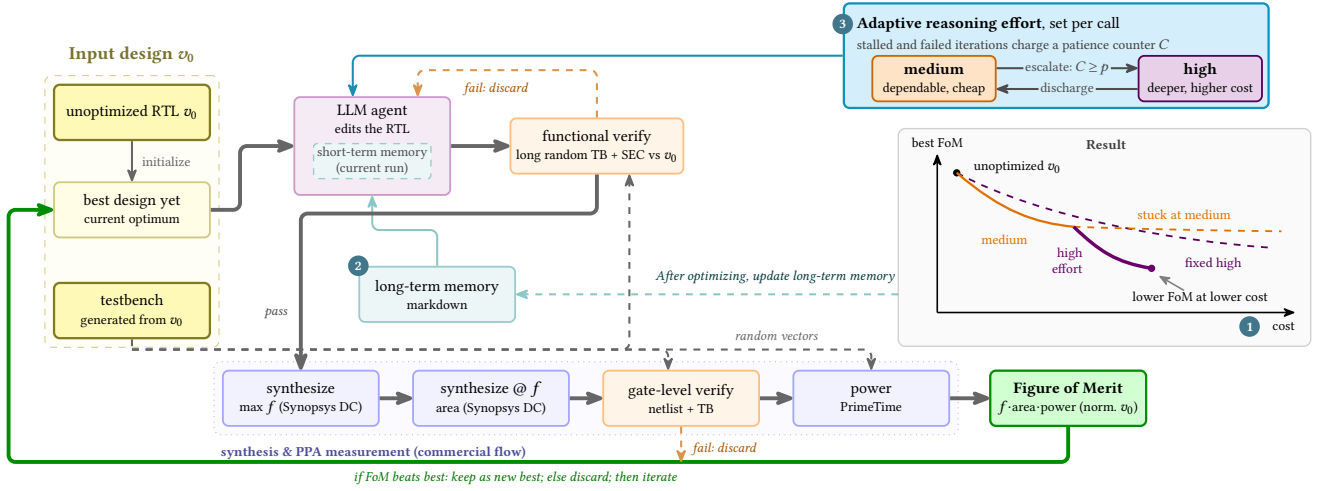

{This missing cost metric also leaves unevaluated where an optimizer's
quality actually comes from.}
{Currently, t}he most capable RTL agents reach a lower FoM by giving the agent
a long-term memory, {a collection of optimization experience
distilled across designs and reused on new ones}~\cite{drrtl,veriagent}.
These agents attribute the quality of the optimizer to how that memory
is constructed, and engineer its content and organization accordingly,
from distilled markdown rule libraries to manager-curated entry
pools~\cite{drrtl,veriagent,memskill,mementoskills}.
{Using a cost-controlled comparison, we find this attribution does
not hold at equal cost: an engineered memory brings no dependable
gain over a plain concatenation of the same experience, and both bring little gain over running with no long-term memory.}

{We instead turn to the reasoning effort the LLM is
allowed to perform per call,} a per-call setting, typically low, medium, or high,
that determines how much hidden reasoning the LLM performs
before it emits an edit, at a higher token cost for a higher
level~\cite{reasonbudget}.
Prior RTL agents nonetheless hold it at a single
value~\cite{drrtl,veriagent,revolution,cktevo,poet}, while a
fixed high effort overthinks the iterations a cheap edit would have
solved and a fixed low effort underthinks the ones that
genuinely need deeper reasoning~\cite{reasonbudget}.
We find that adapting the effort during the run, raising it only on
the iterations where cheaper reasoning has stalled, improves the
average FoM reached at a given cost, as we show in Sec.~\ref{sec:adaptive}.

{Motivated by these observations, we present \sysname{}, an
LLM-agent RTL optimizer that steers the reasoning effort adaptively
across the iterations of a run, accounting for the dollar cost
of every call,} illustrated in Fig.~\ref{fig:teaser}.
This work makes the following contributions.

\begin{itemize}

\item {\textbf{\cbadge{1}~Normalized per-call cost enables fair
optimizer comparison (Sec.~\ref{sec:cost}).}}
{We propose a unified metric that relates each LLM call's
normalized dollar cost to the FoM it reaches, making optimization
quality and cost jointly comparable across effort levels and
optimizers. This metric reveals that \sysname{} reaches a lower FoM
than the state-of-the-art (SotA) Dr.\,RTL~\cite{drrtl} at only 35\% of its cost.}

\item {\textbf{\cbadge{2}~Long-term-memory construction matters
less than assumed (Sec.~\ref{sec:memories}).}}
{We compare an engineered memory that applies the structuring,
filtering, and abstraction operations of prior
work~\cite{drrtl,veriagent,memskill,mementoskills} against a plain
concatenation of the same experience and against no long-term memory.
The engineered construction brings little benefit over the plain one
for both Opus 4.6 and GPT-5.5,
as we measure at equal cost.}

\item {\textbf{\cbadge{3}~Adaptive per-call reasoning effort
improves optimization quality (Sec.~\ref{sec:sched}).}}
{To allocate the reasoning effort efficiently, we implement
adaptive reasoning-effort steering in \sysname{} via a patience counter,
fit once on 21 training designs, that escalates to deeper reasoning
only once progress at a lower effort stalls. On three held-out test
designs, this policy lowers the FoM by 22--28\% with Opus 4.6 and by 18--34\%
with GPT-5.5, where the best fixed effort reaches 20--24\% and 19--29\%,
respectively, at equal normalized cost.}

\end{itemize}

{The central result is that where and when an LLM agent spends
its reasoning budget, rather than how its memory is written, sets the
FoM per dollar it reaches.}
The adaptive policy ends at a lower FoM than any fixed effort level
in five of the six cases of three test designs and
two models (Sec.~\ref{sec:adaptive}).
{\sysname{} further closes on average 91\% of the gap from an
LLM-drafted microscaling~(MX) multiply-accumulate unit to its
hand-optimized counterpart~\cite{mxmac}, and its best run closes the gap entirely, ending even lower than
the counterpart optimized by \sysname{} itself (Sec.~\ref{sec:mx}).}
{On \texttt{controller}, \sysname{} reaches a mean FoM of 0.731
where REvolution~\cite{revolution} and Dr.\,RTL~\cite{drrtl}, scored with the
same FoM, reach
0.977 and 0.771, while spending 35\% of Dr.\,RTL's
cost for the
same number of design iterations (Sec.~\ref{sec:sota}).}

\section{Background and Motivation}
\label{sec:background}

LLM agents entered RTL design one objective at a time: first checking a
design, then generating it, and now optimizing it~\cite{surveyeda}. Early agents
drafted SystemVerilog assertions and testbenches from a
specification~\cite{assertllm,assertionbench,llm4dv}, and attention then moved to
generating correct RTL, where modern LLMs now score so highly on the
VerilogEval and RTLLM benchmarks that the task is considered close to
solved~\cite{verilogeval,rtllm,rtlcoder,verilogcoder,veriagent}. Yet these
designs are often inefficient: they pass their tests while wasting area and
power~\cite{pluto}. The objective accordingly shifted to the quality
of the hardware, and the RTL-OPT benchmark shows that a weak synthesizer can credit the agent
for a rewrite that a commercial tool would make on its own~\cite{rtlopt}.

Iterating on measured feedback improves this hardware quality
further~\cite{verilogcoder,rtlfixer,revolution}, but the first agents that do so keep what
they learn only for the duration of a run. REvolution and POET evolve a population
of candidate designs scored on power, area, and timing~\cite{revolution,poet}, and
CktEvo keeps an archive of its best candidates~\cite{cktevo}. In all three the
stored candidates are the memory itself: no guidance is distilled from them, the
population or archive is discarded when the run ends, and every new design is
optimized from scratch~\cite{revolution,poet,cktevo}.

A second group of agents carries its experience across designs, as distilled
guidance rather than stored candidates. VeriAgent, in the generation setting, keeps
a pool of structured nodes, each a trigger condition paired with natural-language
guidance, which a dedicated manager agent inserts, refines, or discards after every
run~\cite{veriagent}. Dr.\,RTL distills a markdown library of optimization
strategies, including negative ``avoid'' strategies, and reuses it on designs held
out from its construction~\cite{drrtl}. The same deliberate construction appears in
the general-agent literature, where skill libraries transfer best when aligned with
the target domain~\cite{memskill,mementoskills}. Table~\ref{tab:related} compares
the two groups to \sysname{} by the form of their memory and how long it lives. Of
the five systems, only REvolution and Dr.\,RTL release their
code~\cite{revolution,drrtl}, and they are therefore the two we compare against on
our own flow.

\begin{table}[t]
\centering
\caption{Memory-carrying LLM agents for RTL optimization compared to \sysname{}. }%
\label{tab:related}
\setlength{\tabcolsep}{2.6pt}
\renewcommand{\arraystretch}{1.15}
\footnotesize
\begin{tabular}{@{}lccccccc@{}}
\toprule
 & Task & Mem. & Cross- & Power & Cost & Effort & Open \\
 &      & form & design & in FoM & axis & adapt. & source \\
\midrule
REvolution~\cite{revolution} & gen. & pop.     & no  & yes & no & no & yes\\
POET~\cite{poet}             & opt. & pop.     & no  & yes & no & no & no\\
CktEvo~\cite{cktevo}         & opt. & archive  & no  & no  & no & no & no\\
VeriAgent~\cite{veriagent}   & gen. & nodes    & yes & yes & no & no & no\\
Dr.\,RTL~\cite{drrtl}        & opt. & markdown & yes & no  & no\textsuperscript{\dag} & no & yes\\
\sysname{} (ours)            & opt. & markdown & yes & yes & yes & yes & yes\textsuperscript{*}\\
\bottomrule
\end{tabular}

{\raggedright\scriptsize \dag~Reports one aggregate dollar figure for the complete study~\cite{drrtl}.\par}
{\raggedright\scriptsize *Our code is available at \url{https://github.com/KULeuven-MICAS/ARES}.\par}
\end{table}

No agent of either group reports what an LLM call costs or adapts how much
reasoning it receives, as the two rightmost columns of Table~\ref{tab:related} show.
Outside RTL design this is a known inefficiency: a fixed inference-time budget
wastes reasoning on simple problems and cuts it short on hard
ones~\cite{reasonbudget}. %
{Because} the payoff of extra test-time compute varies
with the difficulty of the prompt, compute-optimal scaling allocates that compute
adaptively per prompt~\cite{computeoptimal}. Cost-aware methods apply the same idea
across models and calls, escalating to a costlier model, routing a call, or pruning
a reasoning step only when the expected value justifies the
cost~\cite{frugalgpt,corl,bavt}. All of them adapt the budget of a single, isolated
task, whereas an optimization run is a sequence of iterations where the
difficulty of finding the next improvement changes as the design becomes more optimized. \sysname{} therefore computes the normalized cost of
every call and steers the reasoning effort across the iterations of one run,
escalating on stalled progress through a counter fit once on the training designs.

\section{The \sysname{} Optimizer}
\label{sec:method}

\sysname{} optimizes designs by repeatedly editing its RTL,
scoring candidates with the FoM, and keeping the best version, while adapting
the reasoning effort it spends on each edit.
{As Fig.~\ref{fig:teaser} shows, it tracks two quantities throughout a run,
the quality reached and the normalized dollar cost of reaching it
\cbadge{1}~(Sec.~\ref{sec:cost}).}
Each iteration hands the running-best RTL, initialized to the unoptimized input
design $v_0$, to the LLM agent together with the conversation of the run so far
and the long-term markdown memory \cbadge{2}~{(Sec.~\ref{sec:memories})}.
The agent proposes an edit {at} the effort level set by the adaptive policy
\cbadge{3}~{(Sec.~\ref{sec:effort})}. %
The candidate must first prove functional equivalence to $v_0$, on a long random
testbench and by sequential equivalence checking. It is then synthesized on the
commercial flow, re-verified on the netlist, and its measured area, power, and
delay combine into the FoM, {reported next to the computed dollar cost of
the call that produced it}.
A candidate that lowers the FoM becomes the new running best and the next
iteration chains from it. One that fails to verify, to synthesize, or to
improve is discarded, and either outcome updates the patience counter that sets
the effort of the next call.
{An optimization run ends when its cumulative cost reaches a user-set budget, and returns
the running best as the optimized design.}

\subsection{{Price the PPA gain of each LLM call}}
\label{sec:cost}
\label{sec:flow}

\sysname{} takes two inputs{,} a functionally correct but
unoptimized RTL design $v_0$, and a testbench that exercises it, generated
by the optimizer itself when none is supplied.

The FoM is the product of the three post-synthesis PPA components, each
normalized to the unoptimized input $v_0$,
\begin{equation}
\mathrm{FoM} \;=\;
\frac{\mathrm{area}}{\mathrm{area}_0}\cdot
\frac{\mathrm{power}}{\mathrm{power}_0}\cdot
\frac{\mathrm{delay}}{\mathrm{delay}_0}
\label{eq:fom}
\end{equation}
The normalization fixes $\mathrm{FoM}(v_0)=1$, and a lower FoM is better. Because
all three components enter the product, an edit that trades one component for
another is judged on its net effect.

{The FoM records how deep an optimizer reaches, not what reaching that
depth costs, and that difference accumulates into cost because the
effort levels of one model are priced differently and a run spans dozens of
iterations. Call counts and token counts are the two natural measures of that
cost, and neither suffices. A call count treats effort levels of unequal
price as one equivalent iteration, {erasing a real difference in cost.}
A token count does distinguish the calls, but input,
cached-input, reasoning, and output tokens carry different unit prices, and
commercial tools bill in dollars, leaving token counts an indirect quantity
the user must still convert.}

{\sysname{} therefore computes the normalized dollar cost of every LLM call
and reports it next to the FoM, rather than just an aggregate total.}
The call's logged input (in), {cache read (cr), cache write (cw)}, and output token (out) counts ($t_i$)
are weighted by their published per-token OpenRouter prices ($p$)~\cite{openrouter,cache}, and the weighted sum is the call's dollar cost.
Summing the calls of a run gives its cumulative cost. {The internal reasoning is included in the output tokens \cite{reasontokens}.}
\begin{equation}
\text{Cost} \;=\; \sum_{i=1}^{N}\left(
  p_{\mathrm{in}}\,t_{i,\mathrm{in}}
+ p_{\mathrm{cr}}\,t_{i,\mathrm{cr}}
+ p_{\mathrm{cw}}\,t_{i,\mathrm{cw}}
+ p_{\mathrm{out}}\,t_{i,\mathrm{out}}
\right)
\label{eq:cost}
\end{equation}
Every cost axis {in this paper} reports {that} cumulative cost
in \emph{high-calls}{,} the dollars spent divided by the design's mean
computed cost of one high-effort call.
{Referring the cost to the design's own high-effort price keeps comparisons
unaffected by pricing changes, because whatever the underlying model charges, one
high-call is always the price of one high-effort iteration on that design. It is
also why cumulative cost, rather than a cumulative call or token count, is the
abscissa of every figure below.}
This per-call accounting goes beyond the single aggregate of prior
work~\cite{drrtl} and the turn-count proxy of general-agent
studies~\cite{morewithless}.

\subsection{{Revisit the value of memory construction}}
\label{sec:memories}

{{With this accounting}, we test at equal cost the claim
of prior work that cross-design long-term memory construction dictates the quality of the optimizer. That attribution, made from distilled rule
libraries to manager-curated entry pools~\cite{drrtl,veriagent,memskill,%
mementoskills}, rests on comparisons of the final FoM, and need not hold at equal
cost.}

The agent draws on two memories. The short-term memory is the conversation of the
current {run.} Every iteration resumes the same session and re-feeds its
history to the LLM~\cite{morewithless}{, and the record} is
discarded when the run ends. The long-term memory is the {experience
accumulated across designs,} a markdown document distilled from training
designs and injected into the agent's prompt,
{so that a strategy learned on one design can be reused on another.}

{To test if construction decides the outcome, we hold everything
else fixed and vary only how the same experience is written down, across three
memory-mode variants built from one identical pool of raw experience.}
\cbadgeorange{1} The \emph{memoryless} agent lacks a long-term memory{. It}
still learns within a design, but starts each new design from scratch.
\cbadgeorange{2} The \emph{baseline} memory distills each training run into a
list of simple descriptions of the optimizations that worked on that
design and concatenates those lists.
\cbadgeorange{3} The \emph{engineered} memory uses the same experience
{and creates a superset that incorporates all prior works' techniques.}
{In this rule superset,}
{1)} each accepted edit is recorded as a structured (context, action,
result) case~\cite{dsagent,memento},
{2)} ineffective and harmful edits are kept as explicit anti-optimizations
to avoid~\cite{drrtl}, and
{3)} duplicate entries are merged across designs and the remainder ranked
by measured effectiveness~\cite{veriagent,mementoskills,memskill}.
{Unlike prior work's annotations at the structural or skill
level~\cite{drrtl,veriagent,memskill,mementoskills}, \sysname{} further
anonymizes design-specific names, so that a rule matches on the structure of a
problem rather than on its source.}

\begin{figure}[t]
  \centering
  \includegraphics[width=\columnwidth]{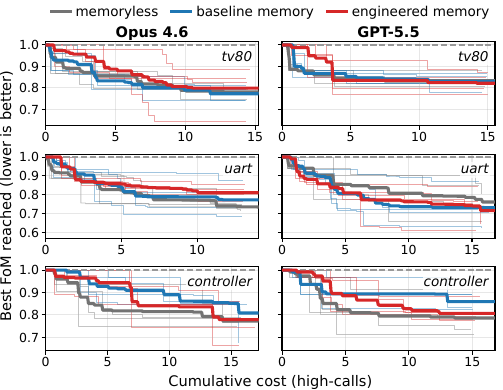}
  \Description{Six step plots in three rows for the test designs tv80, uart, and
  controller, with Opus 4.6 in the left column and GPT-5.5 in the right column.
  Each plot shows the best figure of merit reached, normalized to the input
  design (from 1.0 down to about 0.6, lower is better), against the cumulative
  cost in high-calls (0 to about 16). Thin lines are single runs and thick lines
  are their means, for no long-term memory (gray), the baseline memory (blue),
  and the engineered memory (red). The red mean curve does not stay below the
  blue one: it ends above it in some plots and below it in others.}
  \caption{The three memory constructions of Sec.~\ref{sec:memories} on the
  three test designs{,} best FoM (normalized to $v_0$, lower is better)
  against cumulative cost in high-calls of each model, the unit of Sec.~\ref{sec:cost}. Each
  bold curve is the mean over {four faint individual runs, all using medium reasoning effort.} With both models, the engineered memory
  brings no dependable gain over the baseline concatenation.}
  \label{fig:mem}
\end{figure}

{Starting from identical raw experience, both memory-carrying variants run a full agent optimization per training design and log each edit's FoM effect. Consequently, any FoM difference between them stems solely from how this experience is recorded. As Fig.~\ref{fig:mem} shows (setups in Sec.~\ref{sec:cost} and \ref{sec:setup}), at equal cost the accumulated experience brings little gain over the memoryless agent.}

{Likewise,} the engineered memory (\cbadgeorange{3}) brings no
dependable gain over the {plain concatenation. On all three held-out test
designs and for both models, neither curve is ahead consistently: the engineered memory ends lower
in four of the six cases and higher in two.}
{This means} the structuring, deduplication, and abstraction {that
prior work invests} add nothing on these designs beyond what the baseline
concatenation already delivers.
{What decides the quality of an edit is thus not how much experience is fed
to the LLM, which moves us to examine the optimization process from another
angle.}

\subsection{Steer the per-call reasoning effort}
\label{sec:effort}
\label{sec:sched}

{The other input condition that actually varies between iterations is how
deeply a call is allowed to think. It bears directly on the cost view of
Sec.~\ref{sec:cost}. The price gap between the effort levels is what makes
cost-aware accounting necessary in the first place, and that same gap means the
reasoning effort is not only a cost to account for but a variable the optimizer
can steer. We therefore investigate the per-call reasoning effort. Every}
iteration is one LLM call{, and} the effort, exposed at three
levels, sets how much hidden reasoning that call performs and what it costs.
At low effort the LLM reasons little before emitting its edit, at high
effort it reasons at length, and medium sits between. Changing the level between
iterations leaves the conversation, the long-term memory, and the prompt cache
untouched~\cite{cache}.
Which level an iteration deserves is not known in advance. An opening
iteration with obvious inefficiencies left to remove is well served by a cheap
call, while a run stuck at the same best FoM for many iterations may need
more thinking, and thus a higher reasoning effort, to improve further.

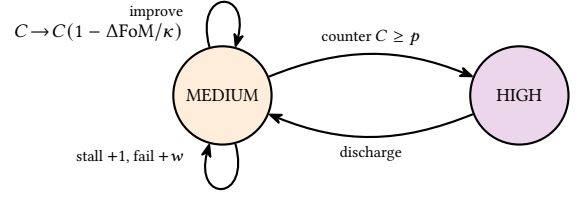
\begin{figure}[t]
\centering
\makebox[\columnwidth][l]{%
\begin{tikzpicture}[
  font=\footnotesize,>={Stealth[round]},
  st/.style={circle,draw,thick,minimum size=13mm,align=center}]
  \node[st,fill=orange!14] (med) {MEDIUM};
  \node[st,fill=violet!16,right=26mm of med] (high) {HIGH};
  \draw[->,thick] (med) to[bend left=22]
      node[above,align=center]{\scriptsize counter $C \ge p$} (high);
  \draw[->,thick] (high) to[bend left=22]
      node[below,align=center]{\scriptsize discharge} (med);
  \draw[->,thick] (med) to[loop above,looseness=6] (med)
      node[above left=6mm and 4mm,align=right]{\scriptsize improve\\$C\!\to\!C(1-\Delta\mathrm{FoM}/\kappa)$};
  \draw[->,thick] (med) to[loop below,looseness=6] (med)
      node[below left=6mm and 4mm,align=right]{\scriptsize stall $+1$, fail $+w$};
\end{tikzpicture}}
\Description{State diagram with two states, MEDIUM and HIGH. An arrow from
MEDIUM to HIGH is labeled with the condition that the counter C reaches p,
and an arrow from HIGH back to MEDIUM is labeled discharge. MEDIUM has two
self-loops: on an improvement the counter C is scaled by one minus the
relative FoM gain divided by kappa, and on a stall it increases by 1, or by w on
a failed edit.}
\caption{The adaptive-effort policy, with constants ($p=3$, $w=2.8$,
$\kappa=0.05$ for Opus 4.6 and $p=4$, $w=4$, $\kappa=0.1$ for GPT-5.5) fit once on the training designs.}
\label{fig:sched}
\end{figure}

\sysname{} therefore starts every design at medium effort, the
dependable middle setting in the fixed-effort comparison of
Sec.~\ref{sec:effortlevels}, and escalates to high effort only once medium has
stalled. We call an iteration \emph{stalled} when its candidate synthesizes and
verifies but does not lower the FoM, and \emph{failed} when it does not even
produce a valid candidate, which could be because the edit does not
compile, synthesize, or verify. A single stalled iteration is a noisy signal,
because even a productive run rejects most of its candidates. \sysname{} instead
accumulates the evidence in a patience counter $C$, updated after every iteration
{as follows.}
\begin{itemize}
\item {A} stalled iteration adds one, $C \to C + 1${.}
\item {A} failed iteration adds a weight $w$, $C \to C + w${.}
\item {An} accepted improvement discharges the counter proportionally to its
      relative FoM gain using a reset scale $\kappa$.
\end{itemize}
Writing the relative gain of an accepted edit as $\Delta\mathrm{FoM} =
(\mathrm{FoM}_{\mathrm{best}} -
\mathrm{FoM}_{\mathrm{new}})/\mathrm{FoM}_{\mathrm{best}}$, the discharge is
$C \to C \cdot \max(0,\, 1 - \Delta\mathrm{FoM}/\kappa)${. A} relative gain
of at least $\kappa$ clears the counter fully, while a marginal gain barely
lowers it. A run that improves marginally therefore
still escalates. When $C$ reaches the patience threshold $p$, the next
iteration runs at high effort and the counter resets, the two-state policy
is shown in Fig.~\ref{fig:sched}.

We fit $p$, $w$, and $\kappa$ jointly, once, on the medium-effort runs of the
training designs of each model. The policy should escalate exactly where staying at medium
would {no longer pay. On} the recorded runs, an escalation is
\emph{warranted} at an iteration where medium produces no accepted improvement
within the next three iterations, and \emph{wasted} where it does. We grid-search the
three constants to maximize the warranted escalations
while minimizing the wasted ones. {This} returns $p=3$, $w=2.8$,
and $\kappa=0.05$ for Opus 4.6 and $p=4$, $w=4$, and $\kappa=0.1$ for
GPT-5.5. That the fitted failure
weight exceeds one is interesting, as the Opus 4.6 training runs {treat} one
failed edit like almost three stalled ones. This can be understood by what
each event reveals{. A} stalled iteration still produces a valid candidate,
and its missing gain may reflect a near-miss of a sound strategy. A failed
iteration produces no valid candidate{,} which could be evidence that the
LLM needs more thinking to properly implement an optimization.

\section{Experimental Setup}
\label{sec:setup}

\begin{figure}[t]
  \centering
  \includegraphics[width=\columnwidth]{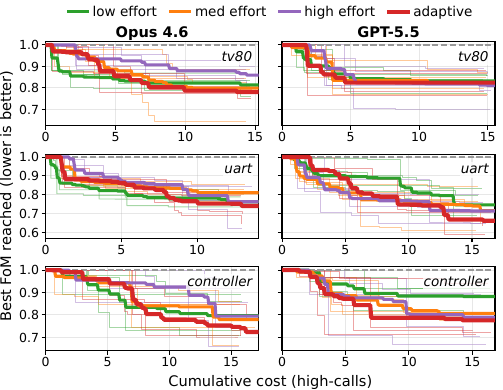}
  \Description{Six step plots in the layout of Figure 2 (tv80, uart, and
  controller in rows; Opus 4.6 left, GPT-5.5 right), with the best normalized
  figure of merit against the cumulative cost in high-calls. The curves compare
  fixed low effort (green), fixed medium effort (orange), fixed high effort
  (purple), and the adaptive policy (red), with thin single runs and thick means.
  The red mean curve ends lowest in five of the six plots; for GPT-5.5 on tv80,
  fixed high effort ends slightly lower.}
  \caption{The three fixed effort levels and the adaptive policy of
  Sec.~\ref{sec:sched} on the three test designs, for Opus 4.6 (left) and
  GPT-5.5 (right), with axes, unit, and
  curve styles as in Fig.~\ref{fig:mem}.}
  \label{fig:effort}
\end{figure}

\textbf{Dataset.} We evaluate on 24 single, self-contained open-source
RTL modules of a few hundred to a few thousand lines: 19 of the 20 designs of
Dr.\,RTL~\cite{drrtl} (its LSTM module is excluded for its inferred memories), an
FFT butterfly and a Huffman decoder from the RTLRewriter
benchmark~\cite{rtlrewriter}, and a CORDIC, a pipelined FFT, and a JPEG DCT from
OpenCores~\cite{opencores}. Three designs spanning
the range of optimizability are held out as the test set: \texttt{tv80}, the
arithmetic-logic unit of a Z80 core; \texttt{uart}, a serial transceiver; and
\texttt{controller}, the control unit of an AES core. The remaining 21 designs form
the training set: one optimization run over each gathers its accepted
optimizations into the long-term memory, and the same runs fit the escalation
constants. Sec.~\ref{sec:mx} additionally optimizes an open-source design
introduced there, an MX multiply-accumulate unit~\cite{mxmac}.
{Because the two long-term memory constructions perform equally, we use the engineered memory in
all experiments of Sec.~\ref{sec:exp}, as we find its structured rules are easier to read for humans.}

\textbf{Testbench and verification.} Verification and power analysis use a
generated testbench of $10^4$ random input vectors from a fixed seed. Every
iteration compares the candidate's outputs, at RTL and on the synthesized netlist,
against those of the unoptimized input design~\cite{poet}. Because random vectors
cannot exercise every structure, equivalence to the input design is additionally
guaranteed by formal sequential equivalence checking at RTL
(JasperGold~\cite{jaspergold}), while the synthesized netlist is still
verified with the testbench.

\textbf{Synthesis flow.} PPA is measured with the commercial Synopsys Design
Compiler and PrimeTime on the open-source Nangate 45\,nm
library~\cite{nangate}. We use a commercial synthesizer because a
weaker open-source one would credit the agent for rewrites a commercial tool
applies on its own~\cite{rtlopt}, and an open-source library so that no
foundry data reaches the LLM. For each design the
synthesizer first finds the shortest achievable clock period, the delay term of the
FoM. The design is then resynthesized at this frequency to obtain area, and
PrimeTime reports power from the switching activity of the testbench vectors
simulated on that netlist~\cite{revolution,veriagent}.

\textbf{LLM.} \sysname{} runs as an agent on the Claude Code command-line
assistant~\cite{claudecode}, the same assistant Dr.\,RTL uses~\cite{drrtl}, on Opus 4.6,
with extended thinking and prompt caching. A call that specifies no effort
level runs at the assistant's default, which is identical to selecting
high~\cite{claudecode}. An effort schedule across iterations must therefore come
from the optimizer. To test transfer to a second model family, \sysname{} also
runs on GPT-5.5~\cite{gpt55} through the Codex command-line
assistant~\cite{codexcli}, at the same effort levels and with the cost computed
in the same way. Because the
LLM cannot be sampled with a fixed seed, each configuration
is repeated multiple times.

\section{Experiments}
\label{sec:exp}

\subsection{No fixed effort level fits the whole run}
\label{sec:effortlevels}

The first experiment varies the fixed per-call reasoning effort of
Sec.~\ref{sec:effort} on the three test designs, and Fig.~\ref{fig:effort}
shows that the reasoning effort changes both what a run costs and
how deep it optimizes.
To ensure a fair comparison, we let every effort level run until it reaches a
similar total cost.{ How well
low effort performs varies with the model: with GPT-5.5 it ends highest on every
design, whereas with Opus 4.6 it does well at equal cost, because its iterations
cost about half as much as medium ones.}

High effort needs more cost than medium to reach the same FoM early in a run in
four of the six cases:
its first successful optimizations land after a cost of 1.8--5.1 high-calls,
where medium cuts its first 5\% of FoM within a cost of 1.1--4.3. The extra reasoning pays off only near the end of
a run: with Opus 4.6, fixed high ends considerably deeper than medium on \texttt{uart} (0.76 against 0.81) but higher on \texttt{tv80} and
\texttt{controller} (0.86 and 0.79 against 0.80 and 0.78){. With
GPT-5.5, however, the difference is smaller: fixed high ends at most 0.017 deeper.}
Medium sits in between: it makes progress early, but it stalls on a plateau that more
spending at medium does not escape, exactly the stall the adaptive policy
exploits.
No fixed level is thus best in every case, and which level is best depends on
both the design and the model.

\subsection{Adaptive effort beats fixed effort}
\label{sec:adaptive}

We next run the adaptive policy of Sec.~\ref{sec:sched}, with the escalation
constants $p$, $w$, and $\kappa$ fixed once on the training designs rather than
tuned per design. The policy starts every design at medium and is just as
cheap early in a run: it cuts its first 5\% of FoM within a cost of 1.1--5.1
high-calls, similar to fixed medium. As
Fig.~\ref{fig:effort} shows, it then overtakes every fixed arm in five of the six
cases: it reaches the final
FoM of fixed high sooner than fixed high itself, at 0.34--0.86$\times$ the cost, and descends beyond it to end deepest. The exception is \texttt{tv80} with GPT-5.5, where fixed high ends
deeper, at 0.81 against 0.82. Across the three designs the adaptive policy lowers the input design's
FoM by 22--28\% with Opus 4.6 and by 18--34\% with GPT-5.5, where the best fixed
arm reaches 20--24\% and 19--29\%, respectively.

{The gain comes from targeting extra effort effectively. Once the fixed medium stalls, spending more on it fails to lower the FoM further, whereas the same cost under the adaptive policy succeeds.}

\subsection{Close the gap to hand-optimized design}
\label{sec:mx}

\begin{figure}[t]
  \centering
  \includegraphics[width=\columnwidth]{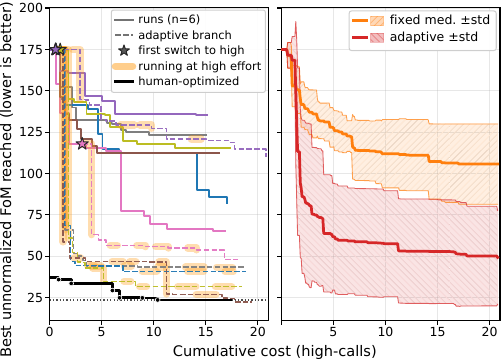}
  \Description{Two step plots of the best unnormalized figure of merit (25 to
  200, lower is better) against the cumulative cost in high-calls (0 to about 21)
  for an LLM-created MX multiply-accumulate unit. The left plot shows six
  fixed-medium-effort runs as solid colored lines. At its first switch to high
  effort, marked by a star, each run branches into a dashed adaptive
  continuation, and orange shading marks the iterations at high effort. Most
  dashed branches drop far below their solid runs. A black step line, the
  human-optimized design under the same optimizer, falls from about 37 to about
  24, and a dotted line marks that level, which the best dashed branch reaches.
  The right plot shows the mean and the standard-deviation band of each arm: the
  fixed-medium mean ends near 105 and the adaptive mean near 50.}
  \caption{An LLM-created MX MAC optimized by \sysname{} on GPT-5.5 toward a human-optimized one (black).
  \textbf{Left:} six fixed-medium runs, each branching to a switch-to-high
  continuation (dashed) at its first stall point, the amber band marking the
  high-effort iterations. \textbf{Right:} the
  same runs averaged (mean$\pm$1\,std bands) for the fixed-medium and
  escalated arms.}
  \label{fig:mx}
\end{figure}

To test \sysname{} where the optimization potential is known to be large, we
select an
open-source MX multiply-accumulate (MAC) unit whose published implementation is
hand-optimized (\texttt{MX\_fp32})~\cite{mxmac}. From its specification and
testbench, Claude Opus 4.6 drafted \texttt{MX\_LLM}, a functionally equivalent MAC,
and \sysname{} optimizes that draft toward the hand-optimized design on the
unnormalized FoM, the absolute area-power-delay product, where the gap between the
two is the improvement to recover. For this MAC, an accuracy check replaces the
equivalence check: on 100,000 random blocks per MX format, every output must stay
within a 0.1\% backward error of an FP64 reference. As Fig.~\ref{fig:mx} shows, \sysname{} lowers
\texttt{MX\_LLM}'s FoM from 174.8 to a mean of 49.3 over six runs,
closing on average 91\% of the gap to \texttt{MX\_fp32}'s published FoM of 36.9. The same optimizer
improves the already hand-optimized \texttt{MX\_fp32} by 36\%, to 23.6, and the best adaptive
\texttt{MX\_LLM} run closes the gap entirely and ends at 22.3, even lower than this optimized
\texttt{MX\_fp32}.

The same six runs measure the adaptive policy's benefit over fixed medium run
by run: each run is branched at its stall point into two continuations, one staying
at medium and one following the adaptive policy, drawn dashed in the left panel of
Fig.~\ref{fig:mx}. The escalated continuation reaches a lower FoM for the same
number of iterations in all six runs. The escalated iterations are more expensive, yet at the same
cumulative cost the adaptive arm still typically sits at a lower FoM, lowering the
mean from 105.7 to 49.3, as the right panel of Fig.~\ref{fig:mx} shows.

The six runs also expose the run-to-run variance of such optimizers: from the
same $v_0$, each fixed-medium run settles onto a plateau of its own, with endpoints scattered from 65.1 to 135.4 (std 24.3). The escalation lowers all six plateaus by 17.2 to 90.3 FoM points, for a mean of 2.8 extra high-calls for the same total number of
iterations. Because a designer pays per run, a dependable result from fixed
medium requires several runs, exactly what the sampling and population strategies
of prior optimizers do~\cite{revolution,poet}. The adaptive policy does not remove this need entirely, due to one run that still
ends at 109.9. A likely explanation for the lower plateaus of the
adaptive policy is that the escalation lands exactly where a run is
stuck, giving the lagging runs the deeper reasoning to leave their local minima.

\subsection{Comparison with SotA optimizers}
\label{sec:sota}

\begin{figure}[t]
  \centering
  \includegraphics[width=\columnwidth]{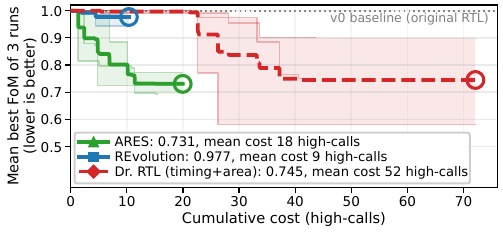}
  \Description{Step plot of the mean best figure of merit over three runs (lower
  is better, 1.0 at the original RTL v0) against the cumulative cost in
  high-calls (0 to 76) on the controller design. Ares (green) falls to 0.731 at a
  mean cost of 18 high-calls, REvolution (blue) stays at 0.977 at a mean cost of
  9, and Dr. RTL (red, dashed, scored with its own timing and area figure of
  merit) reaches 0.745 at a mean cost of 52. Shaded bands show the
  minimum-to-maximum range of the runs, thin lines show single runs, and a ring
  marks the end of each mean curve.}
  \caption{%
  {\sysname{} with adaptive effort, REvolution, and Dr.\,RTL
  optimizing \texttt{controller} from the same $v_0$ on the same flow, three
  runs of 25
  candidates each: mean (thick),
  min-to-max range (band), and single runs (thin). \sysname{} and REvolution use the FoM of
  Eq.~\ref{eq:fom}, with which Dr.\,RTL reaches a mean of 0.771. Dr.\,RTL
  optimizes timing and area only, so its curve uses its own FoM, the normalized
  delay-area product.}}
  \label{fig:sota}
\end{figure}

On a shared commercial tool flow and the FoM of Eq.~\ref{eq:fom}, we compare
\sysname{} against REvolution~\cite{revolution}, the strongest evolutionary LLM-RTL
framework that releases its code, adapted as an optimizer on our flow, and against
the skill-library optimizer Dr.\,RTL~\cite{drrtl} {with its own memory file}. All three optimize
\texttt{controller} from the same $v_0$, three times each. As Fig.~\ref{fig:sota} shows, \sysname{}
reaches a mean FoM of 0.731 at a cost of about 18 high-calls.
REvolution runs its population of five RTL candidates per iteration for five
iterations, as many total designs as \sysname{}'s 25
iterations, and stops at 0.977 at an equivalent cost of 9 high-effort calls on
\texttt{controller}. {We attribute this difference to REvolution lacking both mechanisms} of \sysname{}:
its population stores candidates rather than distilled cross-design experience,
and every call uses the same effort. Dr.\,RTL optimizes timing and
area without power, so its curve
uses its own FoM. Scored with the FoM of Eq.~\ref{eq:fom}, it reaches a mean of
0.771, above the 0.731 of \sysname{}. It
spends 2.8$\times$ \sysname{}'s total over the same 25 iterations,
computed from its transcripts. We attribute the high cost to its multi-agent
pipeline: every iteration runs a critical-path
analysis agent, several parallel rewrite agents, and an evaluation
agent~\cite{drrtl} per proposed design.

\section{Conclusion}
\label{sec:conclusion}

We present \sysname{}, an LLM-agent RTL optimizer that adapts its per-call
reasoning effort with a patience counter and computes the normalized dollar cost
of every LLM call alongside the FoM it reaches.
Its central finding is that the quality and cost of such an
optimizer are set not by how its long-term memory is written but by where its
reasoning is spent. Escalating effort only where medium effort stalls
lowers the input design's FoM by 22--28\% with Opus 4.6 and by 18--34\% with
GPT-5.5, where the best fixed effort reaches 20--24\% and 19--29\%,
respectively. The same optimizer closes on average 91\%
of the gap from an LLM draft to a hand-optimized multiply-accumulate unit and the
full gap in its best run. It also
reaches a lower FoM than prior optimizers at only 35\% of
the cost
of the strongest.

\begin{acks}
This project has been partly funded by the
\grantsponsor{ERC}{European Research Council (ERC)}{https://doi.org/10.13039/501100000781}
under grant agreement No.~\grantnum{ERC}{101088865}, the Flanders AI Research
Program, and long-term structural Methusalem funding by the
\grantsponsor{FlemishGov}{Flemish Government}{https://doi.org/10.13039/501100011878}.
\end{acks}

\bibliographystyle{ACM-Reference-Format}
\bibliography{ares_refs}

\end{document}